\documentclass[aps,pra,10pt,twocolumn,floatfix,nofootinbib]{revtex4-1}

\usepackage{bbm}
\usepackage{amsmath}
\usepackage{amssymb}
\usepackage{graphicx}
\usepackage{amsfonts}
\usepackage{amsthm}

\newtheorem{thm}{Theorem}[section]
\newtheorem{cor}[thm]{Corollary}
\newtheorem{lem}[thm]{Lemma}
\newtheorem{prop}[thm]{Proposition}

\theoremstyle{definition}
\newtheorem{defn}[thm]{Definition}
\newtheorem*{assump1}{Classical assumption}
\newtheorem*{assump2}{Determinism and Reversibility assumption}

\begin{document}

\title{From physical principles to classical Hamiltonian mechanics}
\author{Gabriele Carcassi}
\affiliation{University of Michigan, Ann Arbor, MI 48109}
\email{carcassi@umich.edu}
\date{June 16, 2014}

\begin{abstract}
We derive the Hamiltonian formulation of classical mechanics directly, without reference to Lagrangian mechanics.
We start from the definition of states in terms of labels used to identify them, and show how, under a deterministic and
reversible process, the conservation of the cardinality of the labels leads to Hamilton's equations.\end{abstract}

\maketitle

\section{Introduction}
In physics Hamiltonian mechanics is usually presented as a reformulation of Lagrangian mechanics, which is how it was originally developed. In this work we aim to derive it on its own, starting from simple physical assumptions, without any reference to Lagrangian mechanics.

We'll give simple definitions of states in terms of labels and of determinism/reversibility in terms of bijective maps; define a metric based on the idea that bijective maps preserve the cardinality of labels; obtain Hamilton's equations from the invariance of that metric. In so doing we will be able to answer basic questions such as: what does the geometry of phase space represent? why is evolution continuous? why are conjugate variables special?

No mathematical breakthrough should be expected: the goal, after all, is to derive the \emph{known} framework from a set of \emph{simple} definitions in the most \emph{obvious} way possible. No proof is longer than a couple of paragraphs, so the word \emph{theorem} is avoided in favor of \emph{proposition} and \emph{corollary}. What is surprising how so much can be derived from very few definitions.

\section{States, Labels and Maps}

\begin{defn}\label{statedef}
Suppose we have a physical system to study. We define the set $\mathbbm{C}$ of all physically distinguishable configurations for that system. Each element $\mathbbm{c}$ we call \emph{configuration state}.
\end{defn}

\begin{assump1}\label{classical}
The system is infinitely reducible: it can be thought as composed by two or more similar but smaller systems, each in its own configuration state, which also can be thought as composed by two or more, ad infinitum.
\end{assump1}

\begin{defn}\label{classicalPhaseSpace}
Let $\mathbbm{S}$ be the set of all possible configuration states of the infinitesimal subdivision. We call this set \emph{phase space}. We call each element $\mathbbm{s}$ a \emph{state}.
\end{defn}

\begin{cor}\label{classicalDistribution}
Each classical configuration state $\mathbbm{c} \in \mathbbm{C}$ is a distribution over states: $\mathbbm{c}=\sum\limits_{\mathbbm{s} \in \mathbbm{S}} D(\mathbbm{s}) \mathbbm{s}$, where $D:\mathbbm{S}\rightarrow\mathbb{R}$ measures how much of the system can be found in each $\mathbbm{s}$. The distribution can be visualized as a histogram over the states in phase space.
\end{cor}

Under the classical assumption, we can then limit ourselves to the study of the infinitesimal elements, their states and their properties without loss of generality. To help identify states, we introduce the following concepts.

\begin{defn}\label{label}
We call a \emph{label} a set of states $i\subset\mathbbm{S}$; a \emph{set of labels} a collection of disjoint labels $I | \forall i_1,i_2\in I, i_1\bigcap i_2 = \emptyset$; a \emph{state variable} a set of labels $X$ that covers all of phase space: $\bigcup\limits_{i \in X}i=\mathbbm{S}$. Therefore a state belongs to one and only one label of a state variable.
\end{defn}

\begin{defn}\label{labelsCombine}
Let $I_1$ and $I_2$ be two sets of labels. We can define the \emph{combined set}, $\langle I_1, I_2 \rangle$, whose labels consist of all the non-empty intersections of one label of $I_1$ and one of $I_2$. If all intersections are non-empty, $I_1$ and $I_2$ are said to be \emph{independent}, and we have $n(\langle I_1, I_2 \rangle)=n(I_1)n(I_2)$ where $n: I \rightarrow \mathbbm{R}$ gives the number of labels of each set.
\end{defn}

We now want to study how states and labels evolve in time, under the following assumption.

\begin{assump2}
The system undergoes deterministic (future state identified by the present state) and reversible (past state identified by the present state) evolution.
\end{assump2}

\begin{prop}\label{detrevMap}
Let $\mathbbm{S}$ be the phase space of a system that undergoes deterministic and reversible evolution. There exists a bijective map $f:\mathbbm{S} \leftrightarrow \mathbbm{S}$ between past and future states.
\end{prop}

\begin{cor}\label{detrevDist}
The evolution of a classical configuration state $\mathbbm{c}=\sum D(\mathbbm{s}) \mathbbm{s}$ under a bijective map is given by $\mathbbm{c'}=\sum D'(\mathbbm{s}) \mathbbm{s}=\sum D(f^{-1}(s)) \mathbbm{s}$. The evolution of the fraction of the system in a label $D(i)=\sum\limits_{\mathbbm{S} \in i} D(\mathbbm{s})$ is given by $D'(i)=D(f^{-1}(i))$.
\end{cor}

Mathematically, assuming determinism and reversibility means studying bijective maps. The evolution of a distribution simply moves the elements around: the bars of the histogram move place, but keep the same height.

\begin{cor}\label{labelsCount}
Given a label $i$, the image $f(i)$ is also a label containing the same number of states. Given a set of labels $I$, the image $f(I)$ is also a set of labels containing the same number of labels. Given a state variable $X$, the image $f(X)$ is also a state variable. Given two independent sets of labels $I_1$ and $I_2$, the images $f(I_1)$ and $f(I_2)$ are also independent. Therefore $n(f(\langle I_1, I_2 \rangle))=n(f(I_1))n(f(I_2))=n(I_1)n(I_2)=n(\langle I_1, I_2 \rangle)$
\end{cor}

Bijective maps preserve the number of labels as they provide one-to-one association between future and past. And they do so for each independent set of labels. These simple results, properly generalized to the continuous case, will give us Hamiltonian flow.

\section{Numeric labels}

\textbf{Discrete labels over $\mathbb{R}$}. Numeric labels are often used in physics. For integers, there will be a one-to-one map between $z\in\mathbbm{Z}$ and $i\in I$. Therefore we can use $I$ to both refer to the set of labels and the set of numbers they correspond to. Moreover, from now on we will assume that each numeric label $i$ is enough to identify one and only one state $\mathbbm{s}_i$. This allows us to define the map on the numeric labels directly.

For real numbers, each label corresponds to an interval according to the following:

\begin{defn}\label{discreteLabelDef}
Consider a continuous numeric range. We divide the full range into contiguous cells. Let $I$ be the set of cells. For each cell we have a center value $x: I \mapsto \mathbb{R}$ and a width $w: I \mapsto \mathbb{R}$. $I$ is a set of \emph{discrete labels over a continuous range}.
\end{defn}

\begin{prop}\label{discreteLabelDist}
The configuration state over a set of discrete labels becomes $\mathbbm{c}=\sum D(i) \mathbbm{s}_i$. The fraction of the system in a label is given by $D(i)=\rho(i) w(i)$ where $\rho(i)\equiv D(i) / w(i)$ is the density of the distribution for the cell. In physics terms, the distribution can be visualized as a histogram where $w(i)$, $\rho(i)$ and $D(i)$ are respectively the width, height and area of each bin.
\end{prop}

\begin{cor}\label{discreteLabelEv}
Let $f: I \leftrightarrow I$ be a bijective map. We have $D'(i) = \rho'(i) w(i) = D(f^{-1}(i)) = \rho(f^{-1}(i)) w(f^{-1}(i))$. $\rho'(i) = \rho(f^{-1}(i)) w(f^{-1}(i)) / w(i)$.
\end{cor}

The area moves from one cell of the histogram to the other. The height needs to be adjusted if the cell is of a different size.

\begin{defn}\label{discreteLabelHomogeneous}
A set $I$ of discrete labels over a continuous range is said to be \emph{homogeneous} if $w(i)=k$: the bins are of equal width.
\end{defn}

With homogeneous labels no adjustment is needed, and the range can be used as a measure of the number of labels.\footnote{Since the boundaries between intervals are part of the label, a non-linear transformation will typically transform state variables and their labels from homogeneous to non-homogeneous. Therefore, homogeneity identifies a class of state variables (i.e. a phase space basis) that hold "special status", not for what they represent physically per se, but for their relationship to state definition.}

\textbf{Continuous labels over $\mathbb{R}$}. We now make the bin width arbitrarily small.

\begin{defn}\label{continuousLabels}
A \emph{continuous state variable} $X$ is the continuous limit of a set $I$ of discrete labels over all $\mathbb{R}$.
\end{defn}

To prepare for the limit we define\footnote{This $m(x)$ solves the same problems addressed by the invariant measure $m(x)$ introduced by Jaynes\cite{Jaynes}, which modifies Shannon's differential information entropy definition\cite{Shannon} to be invariant under coordinate transformation.} $m(i)=w(i)/\bar{w}$, where $\bar{w}$ represents the average width of the cells. We increase the number of the cells and reduce $\bar{w}$ while keeping $m(i)$ finite. In the limit we'll have a cell for each value, so we can use $x(i)$ (or simply $x$) instead of $i$ to identify the label. $\rho$ and $m$ will converge to functions defined over $x$. The corresponding configuration state will become $\mathbbm{c}=\int \rho(x) m(x) \mathbbm{s}_x dx$.

\begin{prop}\label{continuousMapping}
Let $f: X \leftrightarrow X$ be a bijective map on a continuous state variable. The mapping must be continuous.
\end{prop}
Assume mapping is discontinuous at point $x$. Consider the cell at $x$ of width $m(x)dx$. The cell would be split into two, so it would not be mapped to one and only one cell. Therefore the mapping must be continuous.

\begin{prop}\label{widthMapping}
Let $f: X \leftrightarrow X$ be a bijective map on a state variable, and $x'=f(x)$. Then $dx' = \frac{m(x')}{m(x)} dx$. If $X$ is homogeneous, then $dx' = dx$; the range gives us a measure of the cardinality of labels and must be conserved.
\end{prop}
The mapping must be done so that the width of the cells is mapped as well, not just the center value. The width of the transported cell $m(x)dx \rightarrow m(x) dx'$ must be equal to the width of the target cell $m(x')dx$, which gives us the first part. If $X$ is homogeneous, $m(x)=m(x')$, which gives us the second part.

Let $\Delta x$ be a finite range of $X$. Before the limit, we have:
\begin{align*}
\Delta x = \frac{\Delta x}{1} = \frac{n(\Delta x) k}{n(1) k} = \frac{n(\Delta x)}{n(1)}
\end{align*}
where $n(\Delta x)$ and $n(1)$ are the number of labels/cells in the $\Delta x$ and unit range, while k is their width. The range can be seen as the ratio between the number of labels in the range and the number of labels in the unit range. That ratio must be conserved by the bijective map and remains well defined during the limit. In this sense, the range can be used as a measure of the cardinality of the labels.

\section{Single degree of freedom}

\begin{defn}\label{sdof}
A \emph{degree of freedom} is a homogeneous two dimensional combined state variable $X=\langle Q, P \rangle$ given by the combined pair\footnote{Having exactly two variables for each d.o.f. is linked to the laws of physics being second order. Its justification is outside of the scope, and will be part of a later work.} of orthogonal state variables $Q$ and $P$. We call these \emph{conjugate variables}.
\end{defn}

\begin{prop}\label{sdofMap}
Let $f: X \leftrightarrow X$ be a bijective map on a degree of freedom. Let $Q'=f(Q)$ and $P'=f(P)$. Then $dq' \wedge dp' = dq \wedge dp$.
\end{prop}

This is the equivalent of \ref{widthMapping} for a two dimensional state variable. The density $\rho(p,q)$ will be defined on cells of infinitesimal area proportional to $dq \wedge dp$. When mapping one cell to another, the infinitesimal area will remain the same. The area in phase space of one degree of freedom can then be used as a measure for the cardinality of the labels. \ref{labelsCount} becomes area conservation\footnote{Under a generic pair of state variables, the conserved quantity related to the cardinality of labels would be $\int\limits_\mathcal{S}\frac{d\hat{q}\wedge d\hat{p}}{m(\hat{q},\hat{p})} $, not the area. Conjugate variables, therefore, simplify the equations by eliminating "apparent" dynamics that would just be the result of the difference in density of non-homogeneous labels.} for conjugate variables.\footnote{P' and Q' remain a homogeneous pair but not, in general, orthogonal.}

\begin{prop}\label{sdofInvariant}
Let $v$ and $w$ be two vectors defined on the tangent space of the manifold identified by two conjugate variables. Let
\begin{align*}
\omega_{\alpha, \beta} = \left[
  \begin{array}{cc}
    0 & 1 \\
    -1 & 0 \\
  \end{array}
\right] \\
\end{align*}
then $v'^{\alpha} \omega_{\alpha, \beta} w'^{\beta}=v^{\alpha} \omega_{\alpha, \beta} w^{\beta}$ under a bijective map.
\end{prop}

Area conservation is equivalent to requiring the invariance of the vector product, which is what $\omega_{\alpha, \beta}$ represents.

\begin{lem}\label{genAntisim}
Let $v$ and $w$ be two vectors. Let $v^{\alpha} \omega_{\alpha, \beta} w^{\alpha}$ be an antisymmetric product conserved under a continuous transformation parameterized by $t$. We can then define a function $H$ such that given $S^{\alpha} \equiv d_{t}x^{\alpha}$ and $S_{\beta} \equiv S^{\alpha} \omega_{\alpha, \beta}$, we have $S_{\alpha} = \partial_{\alpha}H$.
\end{lem}

$S^{\alpha}$ is the vector field that represents how the state variables change. Simply applying the vector transformation rules under continuous transformation we have:
\begin{align*}
v^{\alpha} \omega_{\alpha, \beta} w^{\beta} &= v'^{\alpha} \omega_{\alpha, \beta} w'^{\beta}  \\
&= (v^{\alpha} + \partial_{\gamma} S^{\alpha} dt v^{\gamma}) \omega_{\alpha, \beta} ( w^{\beta} + \partial_{\delta} S^{\beta} w^{\delta} dt) \\
&= v^{\alpha} \omega_{\alpha, \beta} w^{\beta} + (\partial_{\gamma} S^{\alpha} v^{\gamma} \omega_{\alpha, \beta} w^{\beta} \\
 &+ v^{\alpha} \omega_{\alpha, \beta} \partial_{\delta} S^{\beta} w^{\delta}) dt + O(dt^2)
\end{align*}
\begin{align*}
v^{\gamma} w^{\beta} \partial_{\gamma} S_{\beta} - v^{\alpha} w^{\delta} \partial_{\delta} S_{\alpha} = 0
\end{align*}
\begin{align*}
\partial_{\alpha} S_{\beta} - \partial_{\beta} S_{\alpha} &= curl(S_{\alpha}) = 0 \\
S_{\alpha} &= \partial_{\alpha}H
\end{align*}

\begin{prop}\label{sdofHam}
The evolution for a single degree of freedom is given by:
\begin{align*}
d_{t}q &= \partial_{p} H \\
d_{t}p &= - \partial_{q} H
\end{align*}
\end{prop}

Simply expand \ref{genAntisim} with the metric defined in \ref{sdofInvariant}. We recognize Hamilton's equations for one degree of freedom\cite{classical_dynamics}.

\section{Multiple degrees of freedom}

\begin{prop}\label{mdofInvariant}
Let $v$ and $w$ be two vectors defined on the tangent space of the manifold identified by two independent degrees of freedom. Let $\alpha$ and $\beta$ be indexes for the state variables $q^1, p^1, q^2, p^2$. Let
\begin{align*}
\omega_{\alpha, \beta} =  \left[
  \begin{array}{cc}
    1 & 0 \\
    0 & 1 \\
  \end{array}
\right] \otimes \left[
  \begin{array}{cc}
    0 & 1 \\
    -1 & 0 \\
  \end{array}
\right] =
\left[
  \begin{array}{cccc}
    0 & 1 & 0 & 0 \\
    -1 & 0 & 0 & 0 \\
    0 & 0 & 0 & 1 \\
    0 & 0 & -1 & 0 \\
  \end{array}
\right] \\
\end{align*}
then $v'^{\alpha} \omega_{\alpha, \beta} w'^{\beta}=v^{\alpha} \omega_{\alpha, \beta} w^{\beta}$ under a bijective map.
\end{prop}

The independence of the homogeneous state variables corresponds to orthogonality in phase space: from \ref{labelsCombine} the product between the number of labels on each d.o.f. (i.e. the area), must be equal to the number of combines labels (i.e. the hyper-volume), which is true only if the d.o.f are orthogonal in phase space. From \ref{labelsCount}, the mapping will preserve the cardinality of labels, the area\footnote{We assume we are using the same unit across d.o.f.} on each d.o.f, and the independence, orthogonality across d.o.f.\footnote{These statements provide a direct physical interpretation for Gromov's non-squeezing theorem\cite{Gromov,deGosson,Stewart}.} This is equivalent to requiring the conservation of the scalar product across independent degrees of freedom, while still requiring conservation of the vector product within. That leads us to the metric defined by \ref{mdofInvariant}.
The metric generalizes \ref{sdofInvariant} to give us the cardinality of labels defined on the area given by two arbitrary vectors. For an infinitesimal region, this corresponds to $dq^1 \wedge dp^1 + dq^2 \wedge dp^2$, the sum of the projections on the independent planes. Moreover, volume in phase space corresponds to the cardinality of the combined labels (i.e. the states), and is therefore conserved: this is Liouville's theorem.

\begin{prop}\label{mdofHam}
The evolution for multiple degrees of freedom is given by:
\begin{align*}
d_{t}q^i &= \partial_{p^i} H \\
d_{t}p^i &= - \partial_{q^i} H
\end{align*}
\end{prop}

Expand \ref{genAntisim} with the metric defined in \ref{mdofInvariant}. We recognize Hamilton's equations for multiple degrees of freedom\cite{classical_dynamics}.

\section{Time dependence}
So far we have assumed that neither state labeling nor mapping change in time. If they do, we also need to to use time as a label and therefore introduce an extra degree of freedom.

\begin{defn}\label{tdof}
The \emph{temporal degree of freedom} is a homogeneous two dimensional state variable $X=\langle T, E \rangle$ given by the pair of conjugate variables $T$ and $E$. We call \emph{extended phase space} the outer product between phase space and the temporal degree of freedom.
\end{defn}

\begin{prop}\label{tdofMonotonic}
Let $s$ be the parameter of a trajectory in the extended phase space of a deterministic and reversible system. The trajectory must be continuous. There must exist a strictly monotonic function $t(s)$.
\end{prop}

The trajectory has to be continuous in both standard and temporal variables because of \ref{continuousMapping}. Since determinism and reversibility are defined in time, the trajectory must traverse all times once and only once: we must have an invertible mapping between $t$ and $s$, which means we must have a strictly monotonic $t(s)$.

\begin{defn}\label{tdofAntistates}
We call \emph{standard states} those connected by a trajectory where $d_{s}t>0$. We call \emph{anti-states} those connected by a trajectory where $d_{s}t<0$.
\end{defn}

Since $t(s)$ is strictly monotonic, $d_{s}t$ along a trajectory cannot change sign, so we have the division between standard and anti-states. Note that since the parametrization is conventional and can be changed to $s'=-s$, what we call standard and anti-states is also conventional. What is physical and not conventional, though, is that standard and anti-states cannot be connected by deterministic and reversible evolution.

\begin{prop}\label{tdofInvariant}
Let $v$ and $w$ be two vectors defined on the tangent space of the manifold identified by the temporal degree of freedom and one standard degree of freedom. Let $\alpha$ and $\beta$ be indexes for the state variables $t, e, q, p$. Let
\begin{align*}
\omega_{\alpha, \beta} =  \left[
  \begin{array}{cc}
    -1 & 0 \\
    0 & 1 \\
  \end{array}
\right] \otimes \left[
  \begin{array}{cc}
    0 & 1 \\
    -1 & 0 \\
  \end{array}
\right]
= \left[
  \begin{array}{cccc}
    0 & -1 & 0 & 0 \\
    1 & 0 & 0 & 0 \\
    0 & 0 & 0 & 1 \\
    0 & 0 & -1 & 0 \\
  \end{array}
\right] \\
\end{align*}
then $v'^{\alpha} \omega_{\alpha, \beta} w'^{\beta}=v^{\alpha} \omega_{\alpha, \beta} w^{\beta}$ under deterministic and reversible evolution.
\end{prop}

$\langle T, E \rangle$ are not independent from $\langle Q, P \rangle$ as they do not define new states. So they are not necessarily orthogonal in the extended phase space. Looking back at \ref{discreteLabelDef}, cells need to be defined on the plane where $\langle Q, P \rangle$ (maximally) change: this is not the plane of constant $\langle T, E \rangle$ (they are not orthogonal) where $dq \wedge dp$ is defined, but the plane perpendicular to constant $\langle Q, P \rangle$ where $dt \wedge de$ is defined. On that plane we can properly count states and define our invariant.

We have a right triangle-like relationship between the plane where the invariant is defined and its projections on the planes defined by each d.o.f., similar to the multiple d.o.f.:
\begin{align*}
m.d.o.f \;\;\; &dq^1 \wedge dp^1 + dq^2 \wedge dp^2 = k \\
t.d.o.f \;\;\; &dt \wedge de + k = dq \wedge dp \\
\end{align*}
But in the previous case, the right angle was between the two independent d.o.f.. In this case, the right angle is between the invariant and the plane of constant $\langle Q, P \rangle$ where $dt \wedge de$ is defined. We rewrite it as $dq \wedge dp - dt \wedge de = k$. This corresponds to the Minkowski product across d.o.f. and the vector product within. The metric, with a space-like convention, still gives us the cardinality of labels within a degree of freedom.\footnote{Adjusted to avoid double counting.}

\begin{prop}\label{tdofHam}
The evolution for time varying multiple degrees of freedom is given by:
\begin{align*}
d_{s}t &= - \partial_{e} \mathcal{H} \\
d_{s}e &= \partial_{t} \mathcal{H} \\
d_{s}q^i &= \partial_{p^i} \mathcal{H} \\
d_{s}p^i &= - \partial_{q^i} \mathcal{H}
\end{align*}
\end{prop}

Take the metric from \ref{tdofInvariant}, add multiple independent d.o.f as in \ref{mdofInvariant}, use \ref{genAntisim} with the parameter $s$ instead of $t$ and generator $\mathcal{H}$ instead of $H$.

If we set\footnote{We avoided using $p^{n+1}$ as it hides the minus sign from the metric, making it seem that the temporal d.o.f is just another independent d.o.f.} $q^{n+1}=t$ and $p^{n+1}=-e$, we recognise Hamilton equations in the extended phase space\footnote{As in Struckmeier\cite{Struckmeier}, $d_{s}t$ need not be unitary.}\cite{Synge,Lanczos}.

It should not be a surprise that the equations do not mention the speed of light $c$. In fact, nothing says that all $q^i$ represent space or that the laws of motion are invariant in all inertial frames. The only requirement we have is that the areas of each degree of freedom represent the same cardinality for labels.

\begin{prop}\label{tdofConstrain}
The evolution is constrained by $\mathcal{H}=k$.
\end{prop}

Since $\mathcal{H}$ is constant through the evolution, it can serve both as the generating function and as the evolution constraint. By convention, we can set $\mathcal{H}=0$ without loss of generality as changing $\mathcal{H}$ by a constant does not change the equation of motion. This reduces extended phase space to $\mathbb{R}^{2*N + 1}$, the state variables plus time.

\textbf{Example}. We wrap up with an example. Let $\mathcal{H} = mc^2 + ((p^i)^2 - e^2/c^2) / 2m$ = 0.  If we apply \ref{tdofHam} we have:
\begin{align*}
d_{s}t &= e / mc^2 \\
d_{s}e &= 0 \\
d_{s}q^i &= p^i / m \\
d_{s}p^i &= 0
\end{align*}
For standard states (positive $e$), let $s=\tau$:
\begin{align*}
e / c &= m c d_{\tau}t = m U^0 = P^0 \\
p^i &= m d_{\tau}q^i = m U^i = P^i \\
\end{align*}
so we recognize the four-momentum $P^\alpha = [e/c, p^i]$ and $\tau$ proper time. For anti-states (negative $e$), let $s=-\tau$:
\begin{align*}
- e / c &= m c d_{\tau}t = m U^0 = P^0 \\
- p^i &= m d_{\tau}q^i = m U^i = P^i \\
\end{align*}
we end up with a minus sign between the four-momentum and the conjugate variables $P^\alpha = [-e/c, -p^i]$.\footnote{This is consistent with the Dirac field where we have $[i\hbar\partial_t, -i\hbar\partial_{x^i}]$ for generators and $\psi^\dagger\gamma^0[i\hbar\partial_t, -i\hbar\partial_{x^i}]\psi$ for observables, and $\gamma^0$ is $1$ for particles and $-1$ for anti-particles.} $\mathcal{H}$ relates to the rest energy, the parametrization to proper time (its conjugate). Parametrization and time are aligned for standard states and anti-aligned for anti-states.

\section{Conclusion}
By deriving Hamiltonian mechanics from simple definitions of labels, states, determinism and reversibility we have given more direct physical meaning to phase space and its geometric properties, and shown that a good part of the Hamiltonian framework can stand on its own, without Lagrangians. The further emergence of the Minkowski metric and the distinction between standard and anti-states by the inclusion of the temporal d.o.f. is also noteworthy. No mathematical breakthrough is revealed here, yet I am not aware of any work that brings all the pieces of the puzzle together in quite this way: so much derived from so little.

The hope is that, by continuing in this approach, we can shed more light on why the laws of physics are what they are; and show that they are not arbitrary rules, but necessary given few simple assumptions.

\end{document}